# Demonstration of room-temperature continuous-wave operation of InGaAs/AlGaAs quantum well lasers directly grown on on-axis silicon (001)

Chen Jiang,[1,2] Hao Liu,[1,2] Jun Wang,[1,2] Xiao-Min Ren,[1,2,a] Qi Wang,[1,2] Zhuo-Liang Liu,[1,2] Bo-Jie Ma,[1,2] Kai Liu,[1,2] Ren Ren,[1,2] Yi-Dong Zhang,[1,2] Shi-Wei Cai,[1,2] and Yong-Qing Huang[1,2]

[1] State Key Laboratory of Information Photonics and Optical Communications at Beijing University of Posts and Telecommunications, Beijing 100876, China
[2] BUPT-HTGD Joint Laboratory of Quantum Optoelectronics and Bivergentum Theory, Beijing 100876, China

[a] Authors to whom correspondence should be addressed: xmren@bupt.edu.cn

## Abstract

Room-temperature continuous-wave operation of InGaAs/AlGaAs quantum well lasers directly grown on on-axis silicon (001) has been demonstrated. A 420 nm thick GaAs epilayer completely free of antiphase domains was initially grown on the silicon substrate in a metal-organic chemical vapor deposition system and the other epilayers including four sets of five-period strained-layer superlattices and the laser-structural layers were successively grown in a molecular beam epitaxy system. The lasers were prepared as broad-stripe Fabry-Perot ones with a stripe width of 21.5 μm and a cavity length of 1 mm. Typically, the threshold current and the corresponding threshold current density are 186.4 mA and 867 A/cm$^2$, respectively. The lasing wavelength is around 980 nm and the slope efficiency is 0.097 W/A with a single-facet output power of 22.5 mW at an injection current of 400 mA. This advancement makes the silicon-based monolithic optoelectronic integration relevant to quantum well lasers more promising with an enhanced feasibility.

The exponential increase in the size of the internet protocol networks and the urgent demands in information interaction has become a hot issue, both in technology and economy. Optoelectronic integrated circuits (OEICs) fabricated on silicon platforms, combining the merits of mature silicon-based microelectronics technology with high-performance photonic devices, have been regarded as a promising way to break through the limit of the conventional electrical interconnects.[1-3] However, purely silicon-based on-chip light sources remain unavailable due to the indirect bandgap of Si.[4-6] Currently, monolithic III-V semiconductor lasers fabricated on silicon substrates are moving forward by both wafer bonding and heteroepitaxy, but the commercialization requirements for mass production may tilt the balance to the latter.[7] However, in the heteroepitaxy process, material incompatibility between Si and III-V semiconductors usually induces crystal defects including threading dislocations (TDs), antiphase domains (APDs) and microcracks and restricts the laser performance seriously.[8] Actually, great efforts have been made to solve this problem and the suppressions of the last two kinds of defects have been quite successful, but the reduction of the first one, i.e. the TDs, is still not as effective as expected.[9-13]

In recent years, taking the structural advantage of higher tolerance to TDs, significant advancements on silicon-based GaAs-buffered III-V quantum dots (QDs) lasers have been achieved.[14-18] However, in comparison, only some moderate advancements have been made for their quantum well (QW) counterparts although the success of the QW ones with satisfactory performance is so attractive because it would make the transplant of all the currently available GaAs-based or even InP-based III-V optoelectronic technologies onto silicon platforms possible. First of all, by adopting the misoriented silicon substrates, either room-temperature continuous-wave (CW) operation or the room-temperature pulsed operation with a threshold current density as low as 313 A/cm$^2$ have been realized;[19-21] Furthermore, to be compatible with the mature complementary metal-oxide-semiconductor (CMOS) technology, on-axis silicon substrates have also been adopted for the direct GaAs-buffered heteroepitaxial growth, but the simultaneous room-temperature and CW laser operation has not been realized, either being merely room-temperature operation or merely CW operation.[22,23] It should be noted that the room-temperature CW operation of InGaAsP multiple QW lasers fabricated on nano-patterned V-grooved Si (001) substrates have been realized.[24] Nevertheless, due to the complication of such a substrate pre-processing, it would not be a final solution.[25] Therefore, the room-temperature CW operation of silicon-based GaAs-buffered III-V QW lasers directly grown on on-axis Si (001) substrates has not been realized until now.

In this work, by adopting planar on-axis Si (001) substrates along with defect-reduced growth methods, we demonstrate room-temperature continuous-wave CW operation of 980 nm InGaAs/AlGaAs QW lasers. The threshold current and the corresponding threshold current density of the fabricated Fabry-Perot (F-P) lasers are 186.4 mA and 867 A/cm$^2$, respectively. The single-facet output power of 22.5 mW is achieved at an injection current of 400 mA and the slope efficiency is 0.097 W/A.

The whole epitaxial structure of the InGaAs/AlGaAs QW lasers directly grown on the on-axis Si (001) was shown in Fig. 1(a). The actually used substrates were 2-inch ones with their misoriented angles within ±0.5°. At the first stage of the fabrication of the abovementioned structure, the initial 420 nm GaAs epilayer was grown in a metal-organic chemical vapor deposition (MOCVD) system with the III-V precursors of high-purity trimethylgallium (TMGa) and arsine (AsH$_3$). Before loading the substrate into the growth chamber, RCA chemical cleaning was performed to remove the residual contaminants. Then, a hydrogen-annealing pretreatment of the substrate in the growth chamber was done to prevent the APD generation in the following GaAs epilayer growth. Subsequently, the 420 nm GaAs epilayer itself was grown and the growth procedures have been reported in detail previously.[14,26] At the second stage of the fabrication, a prior annealing for the epilayer-surface deoxidation and the growth of the rest parts of the whole structure were conducted in a solid-source molecular beam epitaxy (MBE) system. The detailed procedures were as follows: A 300 nm GaAs buffer was grown first to flatten the surface. Then, four sets of strained-layer superlattices (SLSs) separated by 300 nm GaAs spacing layers were grown as dislocation filter layers (DFLs). Each SLS consisted of five periods of 10 nm $In_{0.166}Ga_{0.834}As$/10 nm GaAs. Subsequently, an 800 nm n-type GaAs contact layer was grown with a Si-doping concentration of $10^{18}/cm^3$. Above this contact layer, a multi-layer core structure was grown. It is actually a ~8 nm $In_{0.16}Ga_{0.84}As$ QW sandwiched by unintentionally doped (UID) $Al_{0.2}Ga_{0.8}As$ waveguide layers and surrounded further by a 1.5 μm Si-doped $Al_{0.4}Ga_{0.6}As$ lower cladding layer and a 1.5 μm Be-doped $Al_{0.4}Ga_{0.6}As$ upper cladding layer. Finally, a 200 nm p-type GaAs contact layer was grown with a Be–doping concentration of $10^{19}/cm^3$. For the sake of comparison, the same laser structure on N-type GaAs substrates was also fabricated.

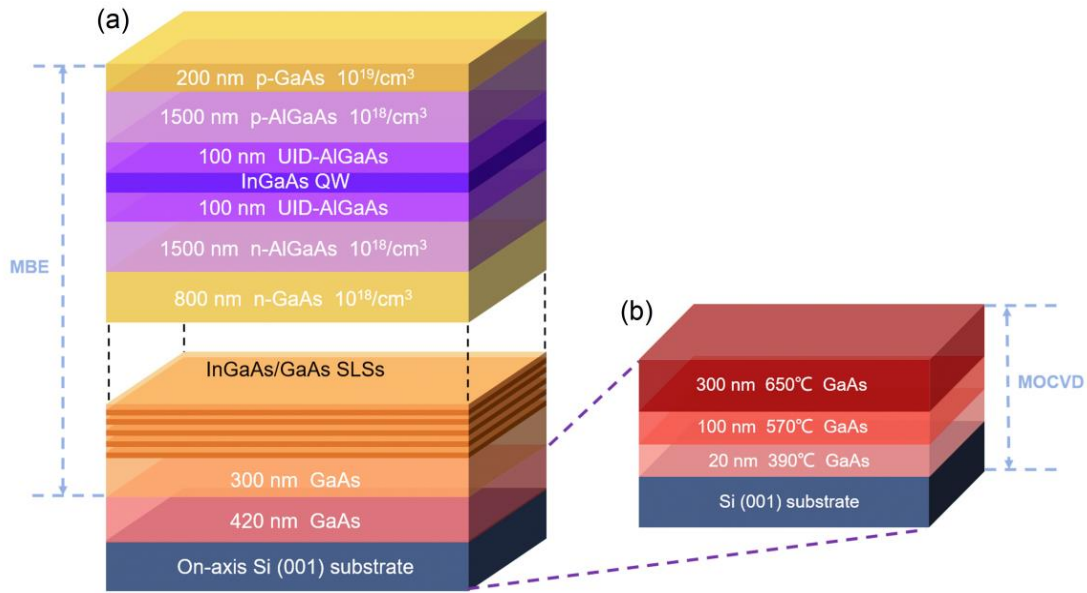

Fig. 1 (a) Schematic of the QW laser directly grown on on-axis Si (001) substrate. (b) Diagram of the initial 420 nm GaAs/Si (001) template grown in MOCVD.

For silicon-based GaAs-buffered InGaAs/AlGaAs QW lasers, the undesired non-radiative recombination in the active layers usually originates from the defect-induced leakage path for minority carriers and the fluctuation (or roughness)-induced generation of delocalized excitons.[27,28] So, to realize desired optical properties, the defect density and roughness of the surface of the epitaxial structure prior to the QW growth should be sufficiently reduced. As the first step, to obtain a smooth epitaxial surface of the GaAs/Si (001) template, the three-step growth method was taken in MOCVD to deposit 420 nm GaAs.[14,26,29] During this process, the APDs, featuring a surface pattern of brink sags in closed loops and further deteriorating the quality of active layers seriously, was completely eliminated by the hydrogen-annealing pretreatment of the silicon substrate.[26] The surface morphology of the 420 nm GaAs/Si (001) template was measured by atomic force microscopy (AFM), as shown in Fig. 2(d). No APD pattern is observed on the surface and the root-mean-square (RMS) roughness is 0.91 nm under the measurement size of 10 μm ×10 μm. As the next step, four sets of SLSs were grown to filter the TDs climbing up from the GaAs/Si interface, as shown in Fig. 2(e). It should be mentioned the proportion of the In composition of the relevant layers in each SLS was all designed as low as 0.166 to alleviate an increasing roughness caused by the accumulated strain from a higher In composition.[30] To estimate the threading dislocation density (TDD) of the n-GaAs surface, a plan-view transmission electron microscopy (PVTEM) measurement was performed. As shown in Fig. 2(c), there are 21 TDs observed in the area of 8.2×5.2 μm$^2$ which corresponds to the TDD of 4.9×10$^7$/cm$^2$. As the last step, the laser-structural layers were grown in an As$_2$-maintained ambient. To further characterize the quality of the InGaAs QW, the photoluminescence (PL) spectra of the GaAs- and Si-based QWs were measured at room temperature (RT, 25°C) under the same excitation power of 48.5 μW, as shown in Fig. 2(a). However, it should be clarified here that the two PL peak intensities are lack of good comparability because the imprecise etching-depth control in the sample preparations led to the different effective excitation powers (the expected etching-depth was ~1.7 μm). As we can see in Fig. 2(a), an obvious peak-wavelength shift happened in comparing the two PL spectra. It would be caused primarily by an unintentional change of the In flux. Meanwhile, the full width at half maximum (FWHM) of the PL spectra on GaAs and Si are 9.2 nm and 14.1 nm, respectively. In addition, the RMS roughness of the top contact-layer surface is 3.81 nm (10×10 μm$^2$), which benefits the subsequent device fabrication. In comparison with the above-mentioned surface roughness of 0.91 nm of the 420 nm GaAs/Si (001) template, the roughness increase should be caused mainly by the doping of the contact layer.

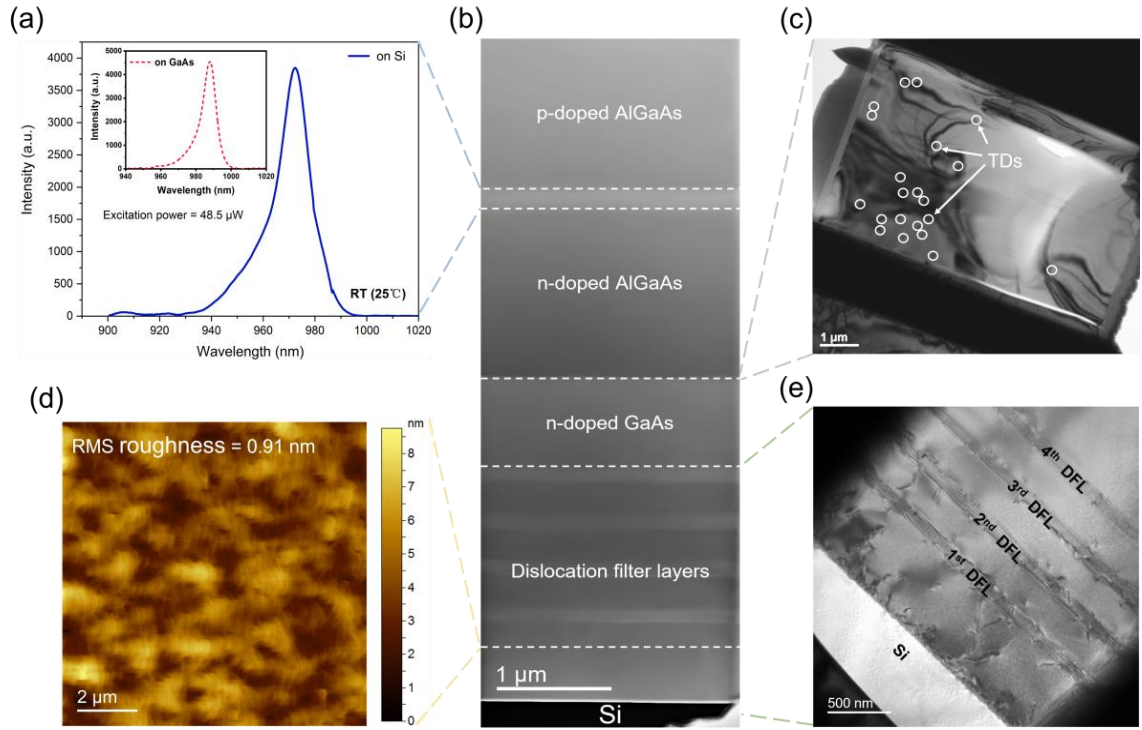

Fig. 2 (a) PL spectra of the laser on Si and the laser on GaAs. (b) Cross-sectional scanning transmission electron microscopy (STEM) image of the whole structure. (c) PVTEM image taken on the surface of the n-GaAs contact layer, the TDD is $4.9×10^7/cm^2$. (d) A typical AFM image of 420 nm GaAs initially grown on the Si substrate in MOCVD with an RMS roughness of 0.91 nm under the scan area of $10×10\ \mu m^2$. (e) Cross-sectional bright-field TEM image of the buffer layers. Four stacks of 5-period InGaAs/GaAs SLSs as DFLs were grown in MBE.

  The schematic of the fabricated laser structure and the cross-sectional scanning electron microscopy (SEM) image of the Si-based QW laser are shown in Fig. 3. The on-chip stripes were made into F-P lasers with $21.5×1000\ \mu m^2$ cavities. Ti/Pt/Au was used as the p-contact metal and AuGe/Ni/Au was used as the n-contact metal. The mirror-like facet, attributing a lower cavity loss, was processed by cleaving without any optical coating. Then, the as-cleaved laser chips were mounted onto Cu heatsinks with C-mount packages and the following tests were carried out under pulsed and CW condition at RT.

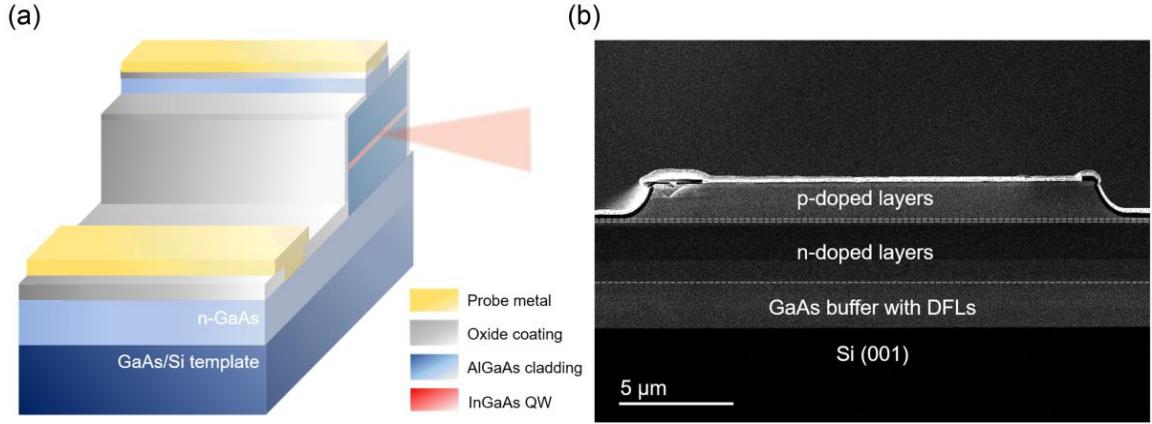

Fig. 3 (a) A 3D schematic of the device structure. (b) Cross-sectional SEM image of as-cleaved facet of the device.

As shown in Fig. 4(a), typical light-current-voltage (L-I-V) properties of broad-stripe GaAs- and Si-based InGaAs/AlGaAs QW lasers were measured at RT (25°C) under pulsed condition (50 μs-pulsed width, 5%-duty cycle). For the laser on the native substrate, the threshold current is 70.5 mA with the corresponding threshold current density of 328 A/cm$^2$. Under the same pulsed condition, the measured threshold current of the laser on the Si substrate is 165.1 mA and the threshold current density is 768 A/cm$^2$. Comparing the Si-based laser with the GaAs-based one, the threshold current density of the former is approximately two times as large as that of the latter, but it is still lower than the values of the QW lasers fabricated on on-axis Si (001) substrates in previous reports.[22,23] From the I-V curves, the differential resistance of the laser on the native substrate is 3.75 Ω, which is 42.8% of 8.76 Ω on the Si substrate. This discrepancy is mainly caused by the long lateral current path, which introduces a larger series resistance in the n-GaAs contact layer. In addition, the slope efficiency is 0.113 W/A for the Si-based laser, and the single-facet output power of 15.5 mW is achieved at an injection current of 300 mA with no sign of power attenuation.

The aforementioned GaAs- and Si-based lasers with the same cavity of 21.5×1000 μm$^2$ were also tested under CW condition at RT (25°C), as shown in Fig. 4(b). The threshold current of the GaAs-based laser is 81.9 mA which corresponds to a threshold current density of 381 A/cm$^2$, and the threshold current of the Si-based one is 186.4 mA, corresponding to a threshold current density of 867 A/cm$^2$. When the injection current reaches 400 mA, the maximum output power and slope efficiency of the GaAs-based laser are 75.6 mW and 0.24 W/A, respectively. For the Si-based one, the maximum output power and slope efficiency are 22.5 mW and 0.097 W/A. It is distinguishing that the performance of the Si-based laser measured under CW condition is poorer than that carried out under pulsed condition. This phenomenon can be illustrated by the self-accumulated heat in the device when operating at a CW injection current. The lasing spectrum of the Si-based laser under CW condition is also presented in Fig. 4(b). When the injection current is 300 mA, the peak wavelength of the spectrum is 980.8 nm. The lifetime of such a Si-based laser under CW condition at RT (25°C) was initially measured as about 90 s. Then, by adopting an optimized device structure, a longer CW lifetime of 12 min at RT (23°C) had been obtained. However, there would still be a long way to go to make the lifetime of these lasers comparable to that of the currently demonstrated Si-based QD lasers and the essential issue to fulfill this target should be dramatically further-reduce the TDD of the heteroepitaxial GaAs virtual substrate.

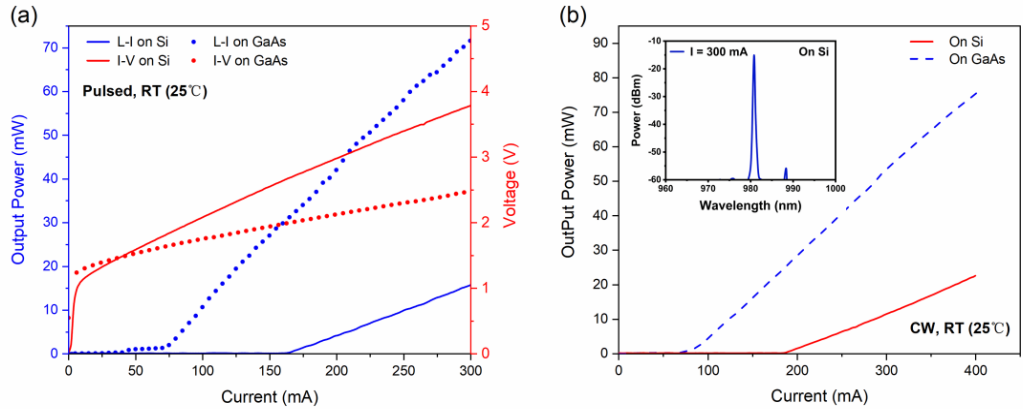

Fig. 4 (a) Typical L-I-V properties of edge-emitting lasers on GaAs and Si substrates. The curves were measured under pulsed condition at RT (25°C). (b) Performances of the GaAs- and Si-based lasers measured under CW condition at RT (25°C). The inset presents the lasing spectrum of the Si-based laser when the injection current is 300 mA.

In conclusion, we have demonstrated continuous-wave electrically pumped 980 nm InGaAs/AlGaAs quantum well lasers directly grown on planar on-axis Si (001). By using a hydrogen-annealing pretreatment of the silicon substrate in an MOCVD system, the APD-free GaAs/Si (001) template with a low surface roughness of 0.91 nm (10×10 μm$^2$) was obtained. Based on this template, we completed the rest laser structures as well as GaAs buffers with DFLs in MBE. The F-P laser with a 21.5-μm stripe width and a 1-mm cavity length exhibited a pulsed lasing at RT with the threshold current of 165.1 mA. The same device operating under CW condition at RT was also measured, the threshold current is 186.4 mA with the corresponding threshold current density of 867 A/cm$^2$. A single-facet output power of 22.5 mW can be achieved at an injection current of 400 mA, and the slope efficiency is 0.097 W/A. This progress indicates the possibility of Si-based QW lasers being applied in commercial products, which promotes further research on the monolithic integration of III-V semiconductor lasers on silicon via heteroepitaxial growth.

## Acknowledgments

The authors would like to thank the support from National Key Research and Development Program of China (No. 2018YFB2200104), Beijing Municipal Science & Technology Commission (No. Z191100004819012), 111 Project of China (No. BP0719012), The Funds for Creative Research Groups of China (No. 62021005), Fund of State Key Laboratory of

## References

[1] S. Y. Siew, B. Li, F. Gao, H. Y. Zheng, W. Zhang, P. Guo, S. W. Xie, A. Song, B. Dong, L. W. Luo, C. Li, X. Luo, and G. Q. Lo, J. Lightwave Technol. **39**, 4374 (2021).

[2] A. H. Safavi-Naeini, D. Van Thourhout, R. Baets, and R. Van Laer, Optica **6**, 213 (2019).

[3] J. C. Norman, D. Jung, Z. Zhang, Y. Wan, S. Liu, C. Shang, R. W. Herrick, W. W. Chow, A. C. Gossard, and J. E. Bowers, IEEE J. Quantum Electron. **55**, 1 (2019).

[4] Z. Wang, A. Abbasi, U. Dave, A. De Groote, S. Kumari, B. Kunert, C. Merckling, M. Pantouvaki, Y. Shi, B. Tian, K. Van Gasse, J. Verbist, R. Wang, W. Xie, J. Zhang, Y. Zhu, J. Bauwelinck, X. Yin, Z. Hens, J. Van Campenhout, B. Kuyken, R. Baets, G. Morthier, D. Van Thourhout, and G. Roelkens, Laser Photonics Rev. **11**, 1700063 (2017).

[5] H. Shen, D.-S. Li, and D.-R. Yang, Acta Phys. Sin. **64**, 204208 (2015).

[6] D. Liang and J. E. Bowers, Nat. Photonics **4**, 511 (2010).

[7] A. Y. L. Sudharsanan Srinivasan, Di Liang, and John E. Bowers, in *Fibre Optic Communication* (2017), pp. 739-797.

[8] J.-S. Park, M. Tang, S. Chen, and H. Liu, Crystals **10**, 1163 (2020).

[9] J. Kwoen, J. Lee, K. Watanabe, and Y. Arakawa, Jpn. J. Appl. Phys. **58**, SBBE07 (2019).

[10] K. Li, J. Yang, Y. Lu, M. Tang, P. Jurczak, Z. Liu, X. Yu, J. S. Park, H. Deng, H. Jia, M. Dang, A. M. Sanchez, R. Beanland, W. Li, X. Han, J. C. Zhang, H. Wang, F. Liu, S. Chen, A. Seeds, P. Smowton, and H. Liu, Adv. Opt. Mater. **8**, 2000970 (2020).

[11] M. Martin, D. Caliste, R. Cipro, R. Alcotte, J. Moeyaert, S. David, F. Bassani, T. Cerba, Y. Bogumilowicz, E. Sanchez, Z. Ye, X. Y. Bao, J. B. Pin, T. Baron, and P. Pochet, Appl. Phys. Lett. **109**, 253103 (2016).

[12] W.-Q. Wei, J.-H. Wang, B. Zhang, J.-Y. Zhang, H.-L. Wang, Q. Feng, H.-X. Xu, T. Wang, and J.-J. Zhang, Appl. Phys. Lett. **113**, 053107 (2018).

[13] T. Ward, A. M. Sánchez, M. Tang, J. Wu, H. Liu, D. J. Dunstan, and R. Beanland, J. Appl. Phys. **116**, 063508 (2014).

[14] J. Wang, Z. Liu, H. Liu, Y. Bai, B. Ma, C. Xiao, C. Jiang, J. Li, H. Wang, Y. Jia, K. Liu, Y. Yang, Q. Wang, Y. Huang, and X. Ren, Opt. Express **30**, 11563 (2022).

[15] C. Shang, E. Hughes, Y. Wan, M. Dumont, R. Koscica, J. Selvidge, R. Herrick, A. C. Gossard, K. Mukherjee, and J. E. Bowers, Optica **8**, 749 (2021).

[16] J. Kwoen, B. Jang, J. Lee, T. Kageyama, K. Watanabe, and Y. Arakawa, Opt. Express **26**, 11568 (2018).

[17] S. Chen, W. Li, J. Wu, Q. Jiang, M. Tang, S. Shutts, S. N. Elliott, A. Sobiesierski, A. J. Seeds, I. Ross, P. M. Smowton, and H. Liu, Nat. Photonics **10**, 307 (2016).

[18] J. Wang, H. Hu, H. Yin, Y. Bai, J. Li, X. Wei, Y. Liu, Y. Huang, X. Ren, and H. Liu, Photonics Res. **6**, 321 (2018).

[19] T. Egawa, H. Tada, Y. Kobayashi, T. Soga, T. Jimbo, and M. Umeno, Appl. Phys. Lett. **57**, 1179 (1990).

[20] M. E. Groenert, C. W. Leitz, A. J. Pitera, V. Yang, H. Lee, R. J. Ram, and E. A. Fitzgerald, J. Appl. Phys. **93**, 362 (2003).

[21] J. Wang, X. Ren, C. Deng, H. Hu, Y. He, Z. Cheng, H. Ma, Q. Wang, Y. Huang, X. Duan, and X. Yan, J.

Lightwave Technol. **33**, 3163 (2015).

[22] X. Huang, Y. Song, T. Masuda, D. Jung, and M. Lee, Electron. Lett. **50**, 1226 (2014).

[23] V. Y. Aleshkin, N. V. Baidus, A. A. Dubinov, A. G. Fefelov, Z. F. Krasilnik, K. E. Kudryavtsev, S. M. Nekorkin, A. V. Novikov, D. A. Pavlov, I. V. Samartsev, E. V. Skorokhodov, M. V. Shaleev, A. A. Sushkov, A. N. Yablonskiy, P. A. Yunin, and D. V. Yurasov, Appl. Phys. Lett. **109**, 061111 (2016).

[24] B. Shi, H. Zhao, L. Wang, B. Song, S. T. Suran Brunelli, and J. Klamkin, Optica **6**, 1507 (2019).

[25] Y. Wan, C. Shang, J. Norman, B. Shi, Q. Li, N. Collins, M. Dumont, K. M. Lau, A. C. Gossard, and J. E. Bowers, IEEE J. Sel. Top. Quantum Electron. **26**, 1 (2020).

[26] W. Chen, J. Wang, L. Zhu, G. Wu, Y. Yang, C. Xiao, J. Li, H. Wang, Y. Jia, Y. Huang, and X. Ren, J. Phys. D: Appl. Phys. **54**, 445102 (2021).

[27] C. Hantschmann, Z. Liu, M. Tang, S. Chen, A. J. Seeds, H. Liu, I. H. White, and R. V. Penty, J. Lightwave Technol. **38**, 4801 (2020).

[28] H. Mehdi, M. Martin, C. Jany, L. Virot, J. M. Hartmann, J. Da Fonseca, J. Moeyaert, P. Gaillard, J. Coignus, C. Leroux, C. Licitra, B. Salem, and T. Baron, AIP Adv. **11**, 085028 (2021).

[29] Y. Wang, Q. Wang, Z. Jia, X. Li, C. Deng, X. Ren, S. Cai, and Y. Huang, J. Vac. Sci. Technol., B: Nanotechnol. Microelectron.: Mater., Process., Meas., Phenom. **31**, 051211 (2013).

[30] M. Tang, S. Chen, J. Wu, Q. Jiang, K. Kennedy, P. Jurczak, M. Liao, R. Beanland, A. Seeds, and H. Liu, IEEE J. Sel. Top. Quantum Electron. **22**, 50 (2016).